\documentclass[prb,showpacs]{revtex4}
\usepackage{bm}% bold math

\def\br{ \bm{r} }
\def\bk{ \bm{k} }
\def\bo{ \bm{0} }
\def\im{ \,\mathrm{Im}\,}
\def\re{ \,\mathrm{Re}\,}
\def\Tr{ \,\mathrm{Tr}\,}
\def\bq{ \bm{q} }

\begin{document}

\title{Does the nuclear spin relaxation rate in superconductors depend on disorder?}

\author{K. V. Samokhin$^{1,2}$ and B. Mitrovi\'c$^{1}$}

\affiliation{$^{1}$ Department of Physics, Brock University,
St.Catharines,
Ontario L2S 3A1, Canada\\
$^{2}$ Commissariat \`a l'Energie Atomique, DSM/DRFMC/SPSMS, 38054
Grenoble, France}

\begin{abstract}
We calculate the relaxation rate of a nuclear spin in $s$-wave
superconductor with nonmagnetic impurities, including the
strong-coupling effects. We show that in a weakly disordered
three-dimensional system the corrections due to disorder are
negligibly small.
\end{abstract}

\pacs{74.25.Nf, 74.20.-z}

\maketitle

\section{Introduction}
\label{sec: intro}

In this article we revisit a long-standing problem about the
effect of nonmagnetic impurities on the nuclear spin relaxation
rate \cite{Slichter90} $R_s(T)$ in isotropic superconductors.  The
first calculation of $R_s$ in a clean superconductor using the
Bardeen-Cooper-Schrieffer (BCS) model was done by Hebel and
Slichter \cite{HS59}. They noticed that the expression for the
relaxation rate is logarithmically divergent at all temperatures
below $T_c$, due to the singularity of the BCS quasiparticle
density of states (DoS) at the gap edge. This singularity survives
even in the presence of scalar disorder (since, according to
Anderson's theorem, the superconducting gap is not affected by
nonmagnetic impurities) and was removed in Ref. \cite{HS59} by
phenomenologically introducing some energy level broadening. A few
years later, Maki and Fulde \cite{MF65} calculated the relaxation
rate in a superconductor with magnetic impurities, in which case
the DoS singularity is smeared and the expression for $R_s$ is
finite at all $T$. It was shown in Ref. \cite{MF65} that the
impurity vertex corrections to the relaxation rate are negligible.
In a parallel development, the Hebel-Slichter arguments about the
importance of the level broadening were put on a quantitative
footing by Fibich \cite{Fibich65}, see also Ref.
\cite{Scalapino69}, who calculated the relaxation rate in the
clean case using the Eliashberg theory of electron-phonon
superconductivity \cite{Elias60} (the so-called strong coupling
model) and found that inelastic scattering indeed removes the
gap-edge singularity and produces a finite relaxation rate.

The effect of scalar impurities in the strong-coupling regime
remained an open problem for another three decades until Choi and
Mele \cite{CM95} presented a calculation of the relaxation rate,
in which they questioned the accuracy of the Maki-Fulde result and
claimed that the impurity vertex corrections are important. As a
result, $R_s$ appreciably increases with nonmagnetic disorder,
especially in the dirty case $\ell\lesssim\xi_0$, where $\ell$ is
the elastic mean free path and $\xi_0$ is the superconducting
coherence length. In the present article we try to resolve this
controversy. The article is organized as follows: In section
\ref{sec: weak coupling}, we calculate the relaxation rate in the
weak-coupling BCS limit, using the exact eigenstates method to
perform the disorder averaging. In section \ref{sec: strong
coupling}, the exact eigenstates method is generalized to the
strong-coupling case. In section \ref{sec: perturbation theory},
we calculate the relaxation rate in the strong-coupling case by
another method, using a standard disorder averaging technique
involving the summation of the ladder impurity diagrams.

\section{Exact eigenstates method: weak coupling}
\label{sec: weak coupling}

Without loss of generality we consider the case of a nuclear spin
$I=1/2$ located at the origin of the crystal lattice. Higher
values of $I$ change only the overall prefactor in the expression
for the relaxation rate \cite{Slichter90}, which drops out of the
ratio of the relaxation rates in the superconducting and the
normal states. We assume, following Hebel and Slichter \cite{HS59}
that, while the resonance is observed in a strong field in the
normal state, the relaxation takes place in a uniform
superconducting state after switching off the field. The
spin-lattice relaxation rate due to the hyperfine contact
interaction of the nucleus with the band electrons is given by
\begin{equation}
\label{T1T general}
    R\equiv\frac{1}{T_1T}=-J^2
    \lim_{\omega_0\to 0}\frac{\im K^R(\omega_0)}{\omega_0},
\end{equation}
where $J$ is the hyperfine coupling constant, $\omega_0$ is the
NMR frequency, and $K^R$ is the retarded correlator of the
transverse components of the electron spin density at the nuclear
site \cite{Mor62,Slichter90}, which is calculated as follows. We
introduce $K(\nu_m)$, with $\nu_m=2\pi mT$, as the Fourier
transform of
\begin{equation}
\label{K_imaginary time}
    K(\tau)=-\langle\langle
    T_{\tau}S_+(\bo,\tau)S_-(\bo,0)\rangle\rangle_{imp}.
\end{equation}
Here $\langle\cdots\rangle$ denotes the quantum statistical
average and $\langle\cdots\rangle_{imp}$ denotes the averaging
over impurity configurations \cite{AGD63,Mahan}. The spin density
operators in the Matsubara representation are given by
$S_\pm(\br,\tau)=e^{H\tau}S_\pm(\br)e^{-H\tau}$, where $H$ is the
electron Hamiltonian, to be defined below, and
\begin{equation}
\label{Spm}
    S_+(\br)=\psi^\dagger_\uparrow(\br)\psi_\downarrow(\br),\quad
    S_-(\br)=\psi^\dagger_\downarrow(\br)\psi_\uparrow(\br).
\end{equation}
We use the units in which $\hbar=k_B=1$, and the spin quantization
axis is chosen along the external magnetic field $\bm{H}$. The
retarded correlator in Eq. (\ref{T1T general}) is obtained by the
analytic continuation:
$K^R(\omega_0)=K(\nu_m)|_{i\nu_m\to\omega_0+i0}$.

The properties of our system in the superconducting state can be
described using a standard field-theoretical formalism in terms of
the normal and anomalous Gor'kov functions \cite{AGD63}, which can
be combined into a $2\times 2$ matrix Green's function
\begin{equation}
\label{Nambu GFs}
    \hat G(\br,\tau;\br',\tau')=\left(
    \begin{array}{cc}
    G(\br,\tau;\br',\tau') & -F(\br,\tau;\br',\tau') \\
    -F^\dagger(\br,\tau;\br',\tau') & -G(\br',\tau';\br,\tau) \\
    \end{array}\right).
\end{equation}
Introducing the two-component Nambu operators \cite{Schr}
\begin{equation}
\label{Nambu_operators}
    \Psi(\br,\tau)=\left(\begin{array}{c}
    \psi_\uparrow(\br,\tau) \\
    \psi^\dagger_\downarrow(\br,\tau)
    \end{array}\right)\quad,\quad
    \bar\Psi(\br,\tau)=(\psi^\dagger_\uparrow(\br,\tau)\quad \psi_\downarrow(\br,\tau)),
\end{equation}
the matrix Green's function can be written in a compact form:
\begin{equation}
\label{Nambu GF def}
    \hat G(\br,\tau;\br',\tau')=-
    \langle T_\tau\Psi(\br,\tau)\bar\Psi(\br',\tau')\rangle.
\end{equation}
The Matsubara spin correlator (\ref{K_imaginary time}) can be
expressed as the impurity average of the product of two matrix
Green's functions:
\begin{equation}
\label{Matsubara correlator}
    K(\nu_m)=\frac{1}2T\sum_n\Tr\langle\hat{G}(\bo,\bo;\omega_n+\nu_m)
    \hat{G}(\bo,\bo;\omega_n)\rangle_{imp},
\end{equation}
where $\omega_n=(2n+1)\pi T$. Note that this representation of the
spin correlator in terms of the $\hat G$'s is not unique, see
section \ref{sec: perturbation theory} below.

We consider a single spin-degenerate electron band in a
three-dimensional (3D) crystal and neglect the spin-orbit
coupling. The electron Hamiltonian is written as $H=H_0+H_{int}$,
where
\begin{equation}
\label{H0}
    H_0=\int d^3\br\;\psi_\alpha^\dagger(\br)h_0
    \psi_\alpha(\br)
\end{equation}
describes noninteracting electrons, with
$\alpha=\uparrow,\downarrow$ being the spin projection, and
\begin{equation}
\label{h0}
    h_0=-\frac{\nabla_{\br}^2}{2m}+U(\br)-\epsilon_F.
\end{equation}
Without loss of generality we assume an isotropic parabolic band.
The impurity potential $U(\br)$ is characterized by the correlator
\begin{equation}
\label{UU-corr}
    \langle U(\br_1)U(\br_2)\rangle_{imp}=\frac{1}{2\pi
    N_F\tau_{el}}\delta(\br_1-\br_2),
\end{equation}
where $N_F$ is the density of states at the Fermi level, and
$\tau_{el}$ is the elastic mean free time.

In a weak-coupling BCS superconductor, the interaction Hamiltonian
which describes $s$-wave singlet pairing has the following form:
\begin{eqnarray}
\label{Hint wc}
    H_{int}=\frac{1}2\int d^3\br\, V_{\alpha\beta,\gamma\delta}
    \psi^\dagger_\alpha(\br)\psi^\dagger_\beta(\br)
    \psi_\gamma(\br)\psi_\delta(\br),
\end{eqnarray}
where
$V_{\alpha\beta,\gamma\delta}=-(\lambda/2)(i\sigma_2)_{\alpha\beta}
(i\sigma_2)^\dagger_{\gamma\delta}$, and $\lambda>0$ is the
coupling constant. Treating the pairing interaction in the
mean-field approximation, the equation of motion for the matrix
Green's function (\ref{Nambu GFs}) can be written in the form
\begin{eqnarray}
\label{Gorkov eqs}
    \left(\begin{array}{cc}
    -\partial_\tau-h_0 & -\Delta(\br) \\
    -\Delta^*(\br) & -\partial_\tau+h_0 \\
    \end{array}\right)
    \hat G(\br,\tau;\br',\tau')=\hat\tau_0\delta(\br-\br')\delta(\tau-\tau'),
\end{eqnarray}
where $\hat\tau_0$ is the unity matrix in the Nambu space, and
$\Delta(\br)=\lambda F(\br,\tau;\br,\tau)$ is the gap function.

The spin correlator (\ref{Matsubara correlator}) will now be
calculated using the exact eigenstates method, see e.g. Ref.
\cite{DeGennes}, which allows one to relate properties of the
system in the superconducting state to those in the normal state.
The key assumption is that the disorder-induced fluctuations of
the gap function can be ignored:
\begin{equation}
\label{average Delta}
    \Delta(\br)\to\Delta=\langle\Delta(\br)\rangle_{imp}.
\end{equation}
According to Anderson's theorem, the average gap is not
renormalized by nonmagnetic impurities, i.e. the value of $\Delta$
at any temperature is the same as in the clean case. We introduce
the exact eigenstates and eigenvalues of the single-particle
Hamiltonian with impurities:
\begin{equation}
\label{ees}
    h_0\varphi_a(\br)=\xi_a\varphi_a(\br),\qquad \int
    d^3\br\;|\varphi_a(\br)|^2=1,
\end{equation}
where, according to the definition (\ref{h0}), the chemical
potential is included in $\xi_a$. Then the Gor'kov equations
(\ref{Gorkov eqs}) can be solved, with the result
\begin{equation}
\label{GF-ee}
    \hat G(\br,\br';\omega_n)=\sum_a\hat
    g_a(\omega_n)\varphi_a(\br)\varphi_a^*(\br'),
\end{equation}
where
\begin{equation}
\label{g_a wc}
    \hat g_a(\omega_n)=-\frac{i\omega_n\hat\tau_0+\xi_a\hat\tau_3
    +\Delta\hat\tau_1}{\omega_n^2+\xi_a^2+\Delta^2}.
\end{equation}
Substituting this in Eq. (\ref{Matsubara correlator}) we obtain,
after the Matsubara frequency summation and the analytic
continuation $i\nu_m\to\omega_0+i0$:
\begin{eqnarray}
\label{KR1}
    \lim_{\omega_0\to 0}\frac{\im K^R(\omega_0)}{\omega_0}
    &=&-\frac{\pi}{2}\Biggl\langle\sum_{a,b}|\varphi_a(\bo)|^2|\varphi_b(\bo)|^2
    \delta(E_a-E_b)\nonumber\\
    &&\times\left(-\frac{\partial f}{\partial E_a}\right)
    \Biggl(1+\frac{\xi_a\xi_b+\Delta^2}{E_aE_b}\Biggr)\Biggr\rangle_{imp},
\end{eqnarray}
where $f(x)=(e^{x/T}+1)^{-1}$ is the Fermi function, and
$E_a=\sqrt{\xi_a^2+\Delta^2}$.

Next, we introduce the local DoS in the normal state:
\begin{equation}
\label{local DOS}
    N(\br,\epsilon)=\sum_a|\varphi_a(\br)|^2\delta(\epsilon-\xi_a),
\end{equation}
perform the disorder averaging in Eq. (\ref{KR1}), and insert the
result in the expression (\ref{T1T general}) for the relaxation
rate:
\begin{equation}
\label{final-result-ee}
    R=J^2N_F^2\int_{-\infty}^\infty d\omega\left(
    -\frac{\partial f}{\partial\omega}\right)\int_{-\infty}^\infty d\epsilon_1
    d\epsilon_2\;\Gamma(\epsilon_1,\epsilon_2;\omega)\,{\cal R}(\epsilon_1-\epsilon_2).
\end{equation}
The superconducting gap enters here only through the function
$\Gamma$, which in the weak-coupling case has the form
$\Gamma=\Gamma_{wc}$, with
\begin{eqnarray}
\label{Gamma wc}
    \Gamma_{wc}(\epsilon_1,\epsilon_2;\omega)&=&\frac{\pi}{2}
    \left(1+\frac{\epsilon_1\epsilon_2+\Delta^2}{\epsilon_1^2+\Delta^2}\right)
    \frac{\sqrt{\epsilon_1^2+\Delta^2}}{|\epsilon_1|}\nonumber\\
    &&\times[\delta(\epsilon_1-\epsilon_2)+\delta(\epsilon_1+\epsilon_2)]
    \delta(\omega-\sqrt{\epsilon_1^2+\Delta^2}),
\end{eqnarray}
while all effects of impurities are contained in the local DoS
correlator
\begin{equation}
\label{cal R def}
    {\cal R}(\Omega)=\frac{1}{N_F^2}
    \langle N(\bo,\epsilon+\Omega)N(\bo,\epsilon)\rangle_{imp}.
\end{equation}

In the normal state, we set $\Delta=0$ in Eq. (\ref{Gamma wc}) and
obtain $R_n=\pi J^2N_F^2{\cal R}(0)$. To calculate the local DoS
correlator, we write
$N(\bo,\epsilon)=-[G^R(\bo,\bo;\epsilon)-G^A(\bo,\bo;\epsilon)]/2\pi
i$, where
\begin{equation}
\label{GRA-def}
    G^{R(A)}(\bo,\bo;\epsilon)=\sum_a\frac{|\varphi_a(\bo)|^2}{\epsilon-\xi_a\pm
    i0}
\end{equation}
are the exact retarded (advanced) Green's functions of the normal
metal. Neglecting the difference between the local and the
Fermi-level densities of states, we replace $\langle
N(\bo,\epsilon)\rangle_{imp}=N_F$, and obtain
\begin{equation}
    {\cal R}(\Omega)=1+\delta{\cal R}(\Omega),
\end{equation}
where
\begin{eqnarray}
    \delta{\cal R}(\Omega)=\frac{1}{2\pi^2}\frac{1}{N_F^2}\re
    [\langle
    G^R(\bo,\bo;\epsilon+\Omega)G^A(\bo,\bo;\epsilon)\rangle_{imp}\nonumber\\
    -\langle G^R(\bo,\bo;\epsilon+\Omega)\rangle_{imp}\langle
    G^A(\bo,\bo;\epsilon)\rangle_{imp}]
\end{eqnarray}
represents the impurity vertex corrections to the product of the
retarded and advanced Green's functions. The terms containing the
averages of two retarded or two advanced Green's functions vanish.
The subsequent steps are standard. Introducing the
disorder-averaged Green's functions
\begin{equation}
    G^{R(A)}(\bk,\epsilon)=\frac{1}{\epsilon-\xi_{\bk}\pm i/2\tau_{el}},
\end{equation}
where $\xi_{\bk}=\bk^2/2m-\epsilon_F$, the impurity ladder
diagrams corresponding to $\delta{\cal R}$ can be summed, with the
following result:
\begin{equation}
\label{delta cal R}
    \delta{\cal R}(\Omega)=\frac{\tau_{el}}{\pi N_F}\re\int\frac{d^3\bq}{(2\pi)^3}
    \frac{\Phi^2(\bq,\Omega)}{1-\Phi(\bq,\Omega)},
\end{equation}
where
\begin{eqnarray}
    \Phi(\bq,\Omega)&=&\frac{1}{2\pi N_F\tau_{el}}\int\frac{d^3\bk}{(2\pi)^3}
    G^R\left(\bk+\frac{\bq}{2},\epsilon+\omega\right)
    G^A\left(\bk-\frac{\bq}{2},\epsilon\right)\nonumber\\
    &=&\left\langle\frac{1}{1-i\Omega\tau_{el}+i\tau_{el}\bm{v}_F(\bk)\bq}\right\rangle_{\hat{\bk}},
\end{eqnarray}
$\bm{v}_F(\bk)$ is the Fermi velocity, and the angular brackets
denote the Fermi-surface averaging. At small $\Omega$ and $\bq$,
$\Phi(\bq,\Omega)\simeq 1+i\Omega\tau_{el}-Dq^2\tau_{el}$, where
$D=v_F^2\tau_{el}/3$ is the diffusion coefficient. Therefore the
impurity vertex corrections lead to a diffusion pole in the
momentum integral on the right-hand side of Eq. (\ref{delta cal
R}). In 3D, due to the convergence of the integral at $q\to 0$,
the dependence of $\delta{\cal R}$ on $\Omega$ is not singular and
can be neglected. On the other hand, $\Phi(\bq,\Omega)$ decays
slowly at $q\to\infty$:
$$
    \Phi(\bq,\Omega)=-\frac{i}{2v_Fq\tau_{el}}\ln\frac{1-i\Omega\tau_{el}+iv_Fq\tau_{el}}{
    1-i\Omega\tau_{el}-iv_Fq\tau_{el}}\stackrel{q\to\infty}{\simeq}
    \frac{\pi}{2v_F\tau_{el}}\frac{1}{q},
$$
for a 3D spherical Fermi surface. Thus it is necessary to
introduce the ultraviolet cutoff of the order of the Fermi
momentum $k_F$ in the integral (\ref{delta cal R}), which gives
the following estimate:
\begin{equation}
    \delta{\cal R}(\Omega)\propto
    \frac{\tau_{el}}{N_F}\int_0^{k_F}q^2dq\,\frac{1}{v_F^2\tau_{el}^2}\frac{1}{q^2}\propto
    \frac{1}{\epsilon_F\tau_{el}},
\end{equation}
and
\begin{equation}
\label{cal R result}
    {\cal R}(\Omega)=
    1+O\left(\frac{1}{\epsilon_F\tau_{el}}\right).
\end{equation}
Although there is some enhancement of the local DoS correlator
and, therefore, of the nuclear spin relaxation rate due to the
diffusive motion of electrons, the magnitude of this effect in a
weakly disordered ($\epsilon_F\tau_{el}\gg 1$) 3D metal turns out
to be negligibly small. Similar conclusions have also been reached
in Ref. \cite{SA94}. Thus we finally arrive at the following
result in the normal state:
\begin{equation}
\label{R n}
    R_n=\pi J^2N_F^2,
\end{equation}
which is known as the Korringa law \cite{Slichter90}.

In the superconducting state, the integral of $\Gamma_{wc}$ on the
right-hand side of Eq. (\ref{final-result-ee}) has a logarithmic
divergency, whose origin can be traced to the square-root
singularity at the gap edge in the BCS density of states. This
singularity is smeared and the divergency is removed in the
presence of magnetic impurities or if the gap is anisotropic
\cite{Tink96}. In this work we focus on the gap smearing due to
the strong-coupling effects in the Eliashberg formalism.

\section{Exact eigenstates method: strong coupling}
\label{sec: strong coupling}

In this section, the exact eigenstates method is generalized to
include the effects of electron-phonon interaction, see also Refs.
\cite{KS76,Bel87}. The Hamiltonian is written as
$H=H_0+H_{ph}+H_{e-ph}$, where $H_0$ describes noninteracting
electrons and is given by Eq. (\ref{H0}),
\begin{equation}
\label{Hph}
    H_{ph}=\sum_{\bq,j}\omega_j(\bq)b^\dagger_{\bq j}b_{\bq j}
\end{equation}
is the Hamiltonian of free phonons, with $j$ labelling the phonon
branches and $\omega_j(\bq)$ being the phonon dispersion (recall
that $\hbar=1$ in our units), and
\begin{equation}
\label{Heph}
    H_{e-ph}=\sum_{\bk_1,\bk_2,j}g_j(\bk_1,\bk_2)(b_{\bq j}+b^\dagger_{-\bq j})
    c^\dagger_{\bk_1\sigma}c_{\bk_2\sigma},
\end{equation}
describes the electron-phonon interaction ($\bq=\bk_1-\bk_2$). The
electron-phonon vertex has the following properties:
$g_j(\bk_1,\bk_2)=g^*_j(\bk_2,\bk_1)$ due to the hermiticity of
$H_{e-ph}$, and $g_j(\bk_1,\bk_2)=g^*_j(-\bk_1,-\bk_2)$ due to
time-reversal symmetry. We assume unit volume and neglect the
effects of impurities on the phonon spectrum. We also neglect the
disorder effect on the electron-phonon vertices, which can be
justified for long-wavelength acoustic phonons by the fact that
electrons move with the lattice, including the impurity atoms, in
order to preserve charge neutrality \cite{Tsun61,Schmid73}.
Although this argument cannot be extended to short-wavelength
phonons, we make the usual assumption that the $g_j$'s are
nonrandom functions.

The matrix Green's function which includes the electron-phonon
interaction but \emph{not} disorder averaging satisfies the
following equation:
\begin{equation}
\label{Dyson-eq}
    \hat G=\hat G_0+\hat G_0\hat\Sigma_{ph}\hat G.
\end{equation}
Here $\hat G_0$ is the matrix Green's function of the disordered
normal metal without phonons, whose coordinate representation can
be obtained by setting $\Delta=0$ in Eq. (\ref{GF-ee}):
\begin{equation}
\label{G0}
    \hat G_0(\br,\br';\omega_n)=-\sum_a
    \frac{i\omega_n\hat\tau_0+\xi_a\hat\tau_3}{\omega_n^2+\xi_a^2}
    \varphi_a(\br)\varphi_a^*(\br'),
\end{equation}
and $\hat\Sigma_{ph}$ is the self-energy due to the
electron-phonon interaction:
\begin{eqnarray}
\label{Sigma-eph}
    \hat\Sigma_{ph}(\bk,\bk';\omega_n)=-T\sum_{n'}\sum_{\bk_1,\bk_1',j}
    \delta_{\bk_1-\bk_1',\bk-\bk'}D_j(\bk-\bk_1,\omega_n-\omega_{n'})\nonumber\\
    \times g_j(\bk,\bk_1)g_j(\bk_1',\bk')\hat\tau_3\hat G(\bk_1,\bk_1';\omega_{n'})\hat\tau_3,
\end{eqnarray}
where
\begin{equation}
\label{D-ph}
    D_j(\bq,\nu_m)=-\frac{2\omega_j(\bq)}{\nu_m^2+\omega^2_j(\bq)}
\end{equation}
is the bare phonon propagator. The contributions to the
self-energy from the diagrams with crossed phonon lines are
neglected based on Migdal's theorem.

The self-energy (\ref{Sigma-eph}) is still a random quantity. To
make progress one has to assume that it can be averaged
independently from the Green's functions, which amounts to
replacing the random self-energy in the equation (\ref{Dyson-eq})
by its disorder average:
\begin{equation}
\label{key-assum}
    \hat\Sigma_{ph}(\bk,\bk';\omega_n)\to\hat\Sigma_{ph}(\bk,\omega_n)=
    \langle\hat\Sigma_{ph}(\bk,\bk';\omega_n)\rangle_{imp}.
\end{equation}
This approximation is justified by the slow spatial variation of
the self-energy compared to that of the electron Green's
functions\cite{KS76}. We use the subscript ``ph'' to emphasize
that $\hat\Sigma_{ph}$ represents only the phononic part of the
full self-energy (the latter contains also the impurity part:
$\hat\Sigma=\hat\Sigma_{ph}+\hat\Sigma_{imp}$). The Coulomb
interaction can be included in a similar fashion. Although there
are impurity vertex corrections to both the electron-phonon and
the Coulomb vertices due to the diffusive motion of electrons, in
a weakly disordered 3D superconductor those are small
\cite{KS76,HM94}. Thus, the average self-energy can be written as
\begin{eqnarray}
\label{average Sigma}
    \hat\Sigma_{ph}(\bk,\omega_n)=T\sum_m\frac{1}{N_F}\sum_{\bk'}\int_0^\infty
    d\Omega\;\alpha^2F(\bk,\bk';\Omega)\frac{2\Omega}{\nu_m^2+\Omega^2}\nonumber\\
    \times\hat\tau_3\hat G(\bk',\omega_n-\nu_m)\hat\tau_3,
\end{eqnarray}
where
\begin{equation}
    \alpha^2F(\bk,\bk';\Omega)=N_F\sum_j|g_j(\bk,\bk')|^2\delta[\Omega-\omega_j(\bk-\bk')],
\end{equation}
and $\hat G(\bk,\omega_n)$ is the disorder-averaged Green's
function of electrons.

For the exact eigenstates method to work we have to neglect the
anisotropy of the electron-phonon interaction, which gives
$\hat\Sigma_{ph}(\bk,\omega_n)=\hat\Sigma_{ph}(\omega_n)$. Similar
to the usual decomposition of the full self-energy,
$\hat\Sigma=i\omega_n(1-Z)\hat\tau_0+\phi\hat\tau_1$ \cite{AM82},
we represent its phononic part in the form
\begin{equation}
\label{Sigma-expansion}
    \hat\Sigma_{ph}(\omega_n)=i\omega_n[1-Z_{ph}(\omega_n)]\hat\tau_0+
    \phi_{ph}(\omega_n)\hat\tau_1,
\end{equation}
where $Z_{ph}$ and $\phi_{ph}$ are real and even functions of the
Matsubara frequency. By analogy with the gap function
$\Delta(\omega_n)\equiv\phi(\omega_n)/Z(\omega_n)$, one can define
$\Delta_{ph}(\omega_n)\equiv\phi_{ph}(\omega_n)/Z_{ph}(\omega_n)$.

Replacing
$\hat\Sigma_{ph}(\bk,\bk',\omega_n)\to\hat\Sigma_{ph}(\omega_n)$
in Eq. (\ref{Dyson-eq}) we find that the electron Green's function
before disorder averaging has the form (\ref{GF-ee}), where $\hat
g_a$ is now given by the following expression:
\begin{equation}
\label{g_a sc}
    \hat g_a(\omega_n)=-\frac{i\omega_nZ_{ph}(\omega_n)\hat\tau_0+\xi_a\hat\tau_3
    +\phi_{ph}(\omega_n)\hat\tau_1}{\omega_n^2Z_{ph}^2(\omega_n)+\xi_a^2
    +\phi_{ph}^2(\omega_n)}.
\end{equation}
Inserting this in Eq. (\ref{Matsubara correlator}), summing over
the Matsubara frequencies and averaging with respect to disorder
one obtains the expression (\ref{final-result-ee}) for the
relaxation rate, with $\Gamma=\Gamma_{sc}$, where
\begin{eqnarray}
\label{Gamma-e1e2}
    &&\Gamma_{sc}(\epsilon_1,\epsilon_2;\omega)=\frac{\pi}{2}
    \Tr[\hat\rho(\epsilon_1,\omega)\hat\rho(\epsilon_2,\omega)],\\
\label{gR}
    &&\hat\rho(\epsilon,\omega)=-\frac{1}{\pi}\im\left.\frac{i\omega_nZ_{ph}(\omega_n)\hat\tau_0
    +\epsilon\hat\tau_3
    +\phi_{ph}(\omega_n)\hat\tau_1}{\omega_n^2Z_{ph}^2(\omega_n)+\epsilon^2
    +\phi_{ph}^2(\omega_n)}\right|_{i\omega_n\to\omega+i0}.
\end{eqnarray}
Using Eq. (\ref{cal R result}), the integrals over $\epsilon_1$
and $\epsilon_2$ can be calculated separately:
$$
    \frac{1}{\pi}\int_{-\infty}^{\infty} d\epsilon\;
    \frac{i\omega_nZ_{ph}(\omega_n)\hat\tau_0+\epsilon\hat\tau_3
    +\phi_{ph}(\omega_n)\hat\tau_1}{\omega_n^2Z_{ph}^2(\omega_n)+\epsilon^2
    +\phi_{ph}^2(\omega_n)}
    =\frac{i\omega_n\hat\tau_0+\Delta(\omega_n)\hat\tau_1}{
    \sqrt{\omega_n^2+\Delta^2(\omega_n)}}.
$$
Here we used the fact that the gap function is not renormalized by
impurities: $\Delta_{ph}(\omega_n)=\Delta(\omega_n)$\cite{AM82}.
Now one can perform the analytic continuation and obtain:
\begin{equation}
    \int_{-\infty}^{\infty} d\epsilon\;\hat\rho(\epsilon,\omega)
    =-\re\frac{\omega\hat\tau_0+\Delta(\omega)\hat\tau_1}{\sqrt{\omega^2-\Delta^2(\omega)}},
\end{equation}
where the branch of the square root is chosen such that its real
part has the same sign as $\omega$, and $\Delta(\omega)$ is the
complex gap function for $\omega$ just above the real frequency
axis \cite{AM82}. Finally, using the normal-state relaxation rate
(\ref{R n}), we obtain:
\begin{eqnarray}
\label{final-result-ee-sc}
    \frac{R_s}{R_n}&=&\int_{-\infty}^\infty
    d\omega\left(-\frac{\partial
    f}{\partial\omega}\right)\nonumber\\
    &&\times\Bigl[\Bigl(
    \re\frac{\omega}{\sqrt{\omega^2-\Delta^2(\omega)}}\Bigr)^2+
    \Bigl(\re\frac{\Delta(\omega)}{\sqrt{\omega^2-\Delta^2(\omega)}}\Bigr)^2\Bigr],
\end{eqnarray}
which coincides with the clean-limit expression derived by Fibich
in Ref. \cite{Fibich65}. This shows the absence of the impurity
effects on the nuclear spin relaxation rate in a weakly disordered
strong-coupling superconductor [up to the terms of the order of
$(\epsilon_F\tau_{el})^{-1}$], which is the main result of this
article.

Note that in general the disorder enters the expression for the
relaxation rate through both the local DoS correlator ${\cal R}$
and $\Gamma_{sc}$, see Eq. (\ref{Gamma-e1e2}). The latter depends
on $Z_{ph}$ and $\phi_{ph}$, which are both renormalized by
disorder. It is the smallness of nontrivial disorder-induced
correlations of the local DoS in 3D, i.e. the fact that ${\cal
R}(\epsilon_1-\epsilon_2)\simeq 1$, that allows one to express the
relaxation rate entirely in terms of the gap function
$\Delta(\omega)$, which is not affected by disorder.

\section{Perturbation theory in the ladder approximation}
\label{sec: perturbation theory}

As an additional check of our result (\ref{final-result-ee-sc}),
we now calculate the spin correlator (\ref{K_imaginary time})
using a direct summation of the impurity vertex corrections in a
standard diagram technique. While some of our intermediate results
look similar to those of Ref. \cite{CM95}, the final conclusion
turns out to be qualitatively very different. We find, in
agreement with the exact eigenstates method of section \ref{sec:
strong coupling}, that the leading order correction to the nuclear
spin relaxation rate resulting from impurity scattering is of the
order of $(\epsilon_F\tau_{el})^{-1}$ in bulk superconductors.

We use the two-component Nambu operators (\ref{Nambu_operators})
in the momentum representation:
\begin{equation}
\label{Nambu_operators_k}
    C_{\bk}(\tau)=\left(\begin{array}{c}
    c_{\bk\uparrow}(\tau) \\
    c^\dagger_{-\bk\downarrow}(\tau)
    \end{array}\right)\quad,\quad
    \bar C_{\bk}(\tau)=(c^\dagger_{\bk\uparrow}(\tau)\quad
    c_{-\bk\downarrow}(\tau)).
\end{equation}
Then the spin-density operators (\ref{Spm}) take the following
form
\begin{eqnarray}
\label{Spm_planewave}
    S_+(\bo,\tau)=\frac{1}2\sum\limits_{\bk,\bk'}
    \bar C_{\bk r}(\tau)(\hat\tau_+)_{rs}\bar C_{\bk' s}(\tau),
                         \nonumber\\
    S_-(\bo,0)=\frac{1}2\sum\limits_{\bk,\bk'}
    C_{\bk r}(0)(\hat\tau_-)_{rs}C_{\bk' s}(0),
\end{eqnarray}
where $r,s=1,2$ are the Nambu indices, and
$\hat\tau_{\pm}=\hat\tau_1\pm i\hat\tau_2$. The Hamiltonian of the
system contains the electron-phonon and the screened Coulomb
interactions as well as the term describing the nonmagnetic
impurity scattering. By applying the Wick theorem to the
correlator (\ref{K_imaginary time}), in which $S_\pm$ are given by
the expressions (\ref{Spm_planewave}), one finds in the clean case
$K(\tau)=(1/2)\sum_{\bk\bk'}\Tr[\hat{G}(\bk,-\tau)(i\hat\tau_2)\hat{G}(\bk',-\tau)
(i\hat\tau_2)]$. In the presence of impurities we obtain the
following form for the disorder-averaged $K(\nu_m)$, which
includes only the second Born approximation impurity ladder
diagrams \cite{AGD63,Mahan}:
\begin{eqnarray}
\label{ladder_Kpm nu m}
    K(\nu_m)=\frac{1}2\sum_{\bk,\bq}T\sum_n\Tr
    [\hat{G}(\bk,\omega_n)(i\hat\tau_2)\hat{G}(\bk+\bq,-(\omega_n+\nu_m))\nonumber\\
    \times\hat{\Gamma}(\bq,\omega_n,\omega_n+\nu_m)],
\end{eqnarray}
where $\hat{G}$ is the matrix Green's function of electrons,
disorder-averaged and fully dressed by all interactions. We assume
a standard isotropic strong-coupling superconductor, for which
\begin{equation}
\label{strong_Nambu GFs omega}
    \hat G(\bk,\omega_n)=-\frac{i\omega_n
    Z(\omega_n)\hat\tau_0+\xi_{\bk}\hat\tau_3+\phi(\omega_n)\hat\tau_1}
    {\omega_n^2Z^2(\omega_n)+\xi_{\bk}^2+\phi^2(\omega_n)},
\end{equation}
with the momentum-independent renormalization function
$Z(\omega_n)$ and the pairing self-energy $\phi(\omega_n)$, which
are even functions of $\omega_n$. Both $Z$ and $\phi$ contain the
effects of impurity scattering, which drop out of the gap function
$\Delta(\omega_n)=\phi(\omega_n)/Z(\omega_n)$ in the isotropic
single-band case \cite{AM82}. The vertex function $\hat{\Gamma}$
satisfies the equation
\begin{eqnarray}
\label{ladder_vertex}
    \hat{\Gamma}(\bq,\omega_n,\omega_n+\nu_m)=i\hat\tau_2+
    \frac{1}{2\pi N_F\tau_{el}}\sum_{\bk}\hat\tau_3\hat{G}(\bk+\bq,-\omega_n-\nu_m)
    \nonumber\\
    \times\hat{\Gamma}(\bq,\omega_n,\omega_n+\nu_m)\hat{G}(\bk,\omega_n)\hat\tau_3.
\end{eqnarray}
Note that Eq. (\ref{ladder_Kpm nu m}) could also be obtained from
the spin correlator (\ref{Matsubara correlator}), using the
identity $(i\hat\tau_2)\hat
G(\bo,\bo,-\omega_n)(i\hat\tau_2)=G(\bo,\bo,\omega_n)$. In the
clean limit we recover from Eqs. (\ref{ladder_Kpm nu m}) and
(\ref{ladder_vertex}) the Fibich's expression \cite{Fibich65} for
the nuclear spin relaxation rate in isotropic single-band
strong-coupling superconductors.

In order to solve Eq. (\ref{ladder_vertex}), one represents the
vertex function as a linear combination of the Pauli matrices in
the Gor'kov-Nambu space:
$\hat\Gamma=\sum_{i=0}^3\Gamma_i\hat\tau_i$, where $\Gamma_i\equiv
\Gamma_i(\bq,\omega_n,\omega_n+\nu_m)=(1/2)\Tr(\hat\tau_i\hat{\Gamma})$,
and finds a set of four coupled algebraic equations:
\begin{equation}
\label{linear_eq} \left.\begin{array}{l}
    \Gamma_0=\displaystyle \sum_j L_{0j}\Gamma_j \nonumber \\
    \Gamma_1=\displaystyle -\sum_j L_{1j}\Gamma_j \nonumber \\
    \Gamma_2=\displaystyle i-\sum_j L_{2j}\Gamma_j \nonumber \\
    \Gamma_3=\displaystyle \sum_j L_{3j}\Gamma_j,
\end{array}\right.
\end{equation}
where $L_{ij}\equiv L_{ij}(\bq,\omega_n,\omega_n+\nu_m)$, with
$i,j=0,1,2,3$, are defined by
\begin{equation}
\label{L_definition}
    L_{ij}=\frac{1}{2\pi N_F\tau_{el}}\frac{1}{2}\sum_{\bk}
    \Tr[\hat\tau_i\hat{G}(\bk+\bq,-\omega_n-\nu_m)
    \hat\tau_j\hat{G}(\bk,\omega_n)].
\end{equation}
One can see that $L_{10}=L_{01}$, $L_{20}=-L_{02}$,
$L_{21}=-L_{12}$, $L_{30}=L_{03}$, $L_{31}=L_{13}$,
$L_{32}=-L_{23}$, so that only ten out of sixteen $L_{ij}$'s in
Eq. (\ref{linear_eq}) have to be computed. The sum over $\bk$ in
Eq. (\ref{L_definition}) is calculated approximately by using
$$
    \sum_{\bk}p(\xi_{\bk},\xi_{\bk+\bq})\simeq\frac{N_F}2\int_{-1}^{1}ds
    \int_{-\infty}^{+\infty} d\xi\; p(\xi,\xi+Q(s)),
$$
assuming that $p$ is decreasing fast enough with $\xi_{\bk}$. Here
$Q(s)=4\epsilon_F(q/2k_F)(q/2k_F+s)$. In this way we find
\begin{eqnarray}
\label{L-results}
    &&L_{00} =
    f_-\left(\frac{\Omega\Omega'+\Phi\Phi'}{DD'}+1\right)\quad,\quad
    L_{01} = if_-\frac{\Omega\Phi'-\Omega'\Phi}{DD'}, \nonumber \\
    &&L_{02} = -if_+\left(\frac{\Phi}{D}+\frac{\Phi'}{D'}\right)\quad,\quad
    L_{03} = if_+\left(\frac{\Omega}{D}+\frac{\Omega'}{D'}\right), \nonumber \\
    &&L_{11} = f_-\left(\frac{\Omega\Omega'+\Phi\Phi'}{DD'}-1\right)\quad,\quad
    L_{12} = f_+\left(\frac{\Omega}{D}-\frac{\Omega'}{D'}\right),  \\
    &&L_{13} = f_+\left(\frac{\Phi}{D}-\frac{\Phi'}{D'}\right)\quad,\quad
    L_{22} = f_-\left(\frac{\Omega\Omega'-\Phi\Phi'}{DD'}-1\right), \nonumber \\
    &&L_{23} = f_-\frac{\Omega\Phi'+\Omega'\Phi}{DD'}\quad,\quad
    L_{33} =
    f_-\left(\frac{\Omega\Omega'-\Phi\Phi'}{DD'}+1\right), \nonumber
\end{eqnarray}
where
\begin{eqnarray*}
    &&\Omega=\omega_n Z(\omega_n),\quad \Phi=\phi(\omega_n),\quad
    D=\sqrt{\Omega^2+\Phi^2},\\
    &&\Omega'=(\omega_n+\nu_m)Z(\omega_n+\nu_m),\quad
    \Phi'=\phi(\omega_n+\nu_m),\quad D'=\sqrt{\Omega'^2+{\Phi'}},
\end{eqnarray*}
and
\begin{eqnarray*}
    f_+ & = & \frac{1}{32\epsilon_F\tau_{el}}
    \frac{1}{x}\ln\frac{x^2(x+1)^2+(D+D')^2/16\epsilon_F^2}{x^2(x-1)^2
    +(D+D')^2/16\epsilon_F^2} \\
    f_- & = & \frac{1}{16\epsilon_F\tau_{el}}
    \frac{1}{x}\left[\tan^{-1}\frac{D+D'}{4\epsilon_Fx(x-1)}
    -\tan^{-1}\frac{D+D'}{4\epsilon_F x(x+1)}\right],
\end{eqnarray*}
with $x=q/2k_F$. We note that the relationship between our
$f_{\pm}$ and $\langle f_{\pm}\rangle$ of Ref. \cite{CM95} is
$\langle f_+\rangle=8if_+$, $\langle f_-\rangle=-8f_-$, with the
additional difference that our $f_{\pm}$ contain
$(D+D')/4\epsilon_F$, instead of $(D+D')/\epsilon_F$.

We solved the system (\ref{linear_eq}) with the $L_{ij}$'s defined
by the expressions (\ref{L-results}), using MAPLE. The result is
\begin{equation}
\label{gamma-solutions} \left.\begin{array}{l}
    \Gamma_0=\displaystyle \frac{f_+}{(1-2f_-)^2+4f_+^2}
    \left(\frac{\Phi}{D}+\frac{\Phi'}{D'}\right),\nonumber \\
    \Gamma_1=\displaystyle -i\frac{f_+}{(1-2f_-)^2+4f_+^2}
    \left(\frac{\Omega}{D}-\frac{\Omega'}{D'}\right), \nonumber \\
    \Gamma_2=\displaystyle i\left[1-\frac{f_--2(f_+^2+f_-^2)}{(1-2f_-)^2+4f_+^2}
    \left(\frac{\Omega\Omega'-\Phi\Phi'}{DD'}-1\right)\right], \nonumber \\
    \Gamma_3=\displaystyle -i\frac{f_--2(f_+^2+f_-^2)}{(1-2f_-)^2+4f_+^2}
    \frac{\Omega\Phi'+\Omega'\Phi}{DD'}.
\end{array}\right.
\end{equation}
Inserting these into the vertex function in Eq. (\ref{ladder_Kpm
nu m}), we finally obtain:
\begin{equation}
\label{ladder_Kpm final}
    K(\nu_m) = -T\sum_{n=-\infty}^{\infty}A_nB_n,
\end{equation}
where
\begin{eqnarray*}
    A_n=\frac{\omega_n(\omega_n+\nu_m)-
    \Delta(\omega_n)\Delta(\omega_n+\nu_m)}{\sqrt{\omega_n^2+
    \Delta^2(\omega_n)}\sqrt{(\omega_n+\nu_m)^2+
    \Delta^2(\omega_n+\nu_m)}}-1,\\
    B_n=2\pi\tau_{el}N_F\sum_{\bf q}\left[f_--
    2\frac{f_+^2}{(1-2f_-)^2+4f_+^2}
    +2\frac{f_-(f_--2(f_+^2
    +f_-^2))} {(1-2f_-)^2+4f_+^2}\right].
\end{eqnarray*}
Clearly, in the limit $\tau_{el}\rightarrow\infty$ only the first
term in the square bracket in $B_n$ survives, and it is possible
to integrate it over $\bq$ analytically, with the result
$\pi^2N_F^2\sqrt{(1+\sqrt{1+\rho^2})/2}$, where
$\rho=(D+D')/\epsilon_F$. For all the terms in the sum over $n$ in
Eq. (\ref{ladder_Kpm final}) for which $A_n$ is nonzero [note that
$\Delta(\omega_n)\simeq$ 0 for $|\omega_n|$ greater than 10 times
the maximum phonon frequency], $\rho$ can be set equal to zero. In
this way one recovers Fibich's formula \cite{Fibich65,Scalapino69}
for the relaxation rate, after the sum over $n$ is performed,
followed by the analytic continuation $i\nu_m\to\omega_0+i0$, in
the limit $\omega_0\to 0$. We note that our $B_n$, containing the
impurity vertex corrections, is different from the expression for
the vertex corrections obtained by Choi and Mele \cite{CM95},
which we have not been able to reproduce. An additional difference
is that we calculate analytically the momentum integrals which
were treated in Ref. \cite{CM95} using some approximation.

Even for a finite $\tau_{el}$ one can replace $\rho\to 0$, which
makes it possible to integrate over $\bq$ the second and third
terms in the expression for $B_n$. Our final result for the
nuclear spin relaxation rate in the superconducting state to the
leading order in $(\epsilon_F\tau_{el})^{-1}$ is
\begin{eqnarray}
\label{strong_T1T final result}
    R_s&=&\pi J^2N_F^2
    \left(1+\frac{3\pi}{16}\frac{1}{\epsilon_F\tau_{el}}\right)
    \int_{-\infty}^{\infty}d\omega
    \left(-\frac{\partial f}{\partial\omega}\right) \nonumber \\
    &&\times\Bigl[\Bigl(
    \re\frac{\omega}{\sqrt{\omega^2-\Delta^2(\omega)}}\Bigr)^2+
    \Bigl(\re\frac{\Delta(\omega)}{\sqrt{\omega^2-\Delta^2(\omega)}}\Bigr)^2\Bigr].
\end{eqnarray}
We see that the impurity vertex corrections turn out to be of the
same order as the diagrams with crossed impurity lines, which we
have neglected, and the ratio $R_s/R_n$ is therefore unaffected by
impurity scattering at $\tau_{el}^{-1}\ll\epsilon_F$. This
condition is much weaker than $\ell\gg\xi_0$, i.e.
$\tau_{el}^{-1}\ll\Delta_0$ ($\Delta_0$ is the superconducting gap
at zero temperature), found in Ref. \cite{CM95}.

\section{Conclusions}

We conclude that the answer to the question in the title is
negative. Using two different techniques, the exact eigenstates
method and the usual diagrammatic perturbation theory in the
ladder approximation, we have shown that the contribution of
nonmagnetic impurities to the nuclear spin relaxation rate in a
bulk superconductor with isotropic pairing is of the order of
$(\epsilon_F\tau_{el})^{-1}$, i.e. very small.

\acknowledgements

This work was supported by the Natural Sciences and Engineering
Research Council (NSERC) of Canada.

\end{document}